\documentclass[a4paper,11pt]{article}

\usepackage{jcappub} 

\usepackage[T1]{fontenc} 
\usepackage{multirow,subcaption}

\title{\boldmath A Collaborative Explanation of Cosmic Ray Spectrum Based on the Gluon Condensation Model}

\author[a]{Jintao Wu,}
\author[a,1]{Jianhong Ruan\note{Corresponding author.}}

\affiliation[a]{Department of Physics, East China Normal University,\\Shanghai 201100, China}

\emailAdd{jin-taowu@qq.com}
\emailAdd{jhruan@phy.ecnu.edu.cn}

\abstract{Based on the Gluon Condensation (GC) model, the relationship between the spectra of electrons, $\gamma$ rays, and neutrinos in cosmic rays can be deduced. It has been found that these particles share the same parameter, $\beta_p$, and have an identical GC threshold values. This paper explores the connection between the second excess spectra of electron and the spectra of gamma rays and neutrinos. According to the observed gamma-ray data, it is suggested that the source LHAASO J2108+5157 might contribute to the second excess of electron.
	
\keywords{ gluon condensation, electron spectra, $\gamma$ spectra, neutrino spectra } }

\begin{document}
\maketitle
\flushbottom

\section{Introduction}
\label{sec:intro}
In recent decades, significant advancements have been made in scientific research through both space-based and ground-based experiments. A new generation of experiments has ushered in an era of high precision measurement of cosmic rays (CR), revealing a range of new phenomena. Notably, collaborative groups such as the Alpha Magnetic Spectrometer (AMS-02) \cite{AMS:2019iwo,aguilar2021alpha}, the Fermi Large Area Telescope (Fermi-LAT) \cite{Fermi-LAT:2017bpc}, and the Dark Matter Particle Explorer (DAMPE) \cite{DAMPE:2017fbg,DAMPE:2019gys} have provided relatively precise experimental data of electron and positron ($e^-+e^+$) as well as  proton. The data cover an energy range from approximately 1 GeV to several TeV.

In these experiments, AMS-02 observed a sharp drop in the spectrum of leptons (electrons and positrons) at an energy about 284 GeV, the significant result being the excesses in electron spectra. The DAMPE experiment showed that at energies as high as about 0.9 TeV, the leptons spectrum exhibits a power-law shape, with a spectral break occurring at around 0.9 TeV \cite{DAMPE:2017fbg}. All spectra display features deviating from a single power law, indicating the presence of new sources of CR electrons. The characteristics of CR spectra may be influenced by the nearby sources, including pulsar wind nebulae (PWNe) \cite{DiMauro:2014iia,Hooper:2017gtd,Bykov:2019mis,Manconi:2020ipm,Evoli:2020ash,Bao:2020ila,Ding:2020wyk}, supernova remnants (SNR) \cite{shen1970pulsars,Atoian:1995ux,Kobayashi:2003kp,di2014interpretation,Fang:2017tvj,Evoli:2020szd,Evoli:2021ugn}, as well as dark matter (DM) particle annihilation or decay \cite{Liu:2017rgs,Huang:2017egk,Coogan:2019uij,Liu:2019iik,Feng:2019rgm,Ge:2020tdh,Chen:2021yde}, and others.

The widely adopted model usually treates the total spectrum as the sum of the background SNR and the local SNR sources. For primary nuclei (protons, He, C, O), observations by AMS-02 have shown significant excesses at around 200 GeV \cite{AMS:2015tnn}. DAMPE confirmed the hardening of the proton spectrum \cite{DAMPE:2017fbg}. This spectral excess has been confirmed in observations by the ATIC and NUCLEON experiments, implying that the excesses of nuclei and electrons may be accelerated by a same local SNR \cite{Mertsch:2010fn,Bernard:2012wt,Serpico:2011wg,Fang:2018qco,Tang:2018wyr}.

The research work \cite{Wu:2023fbw} has proposed a collaborative explanation for the spectra excesses of electron and proton observed by DAMPE and AMS-02. Through the analysis of the electron spectrum from DAMPE, the study suggests a possible existence of a second excess near the upper limit of experimental measurement. 
However, due to limited data and large errors, the magnitude and the energy region of this excess cannot be determined from experiments presently. 
According to  the GC model \cite{Zhu:2008ur,Zhu:2016qif,Zhu:2017ydo,Zhu:2017bvp,Feng:2018wio},  there are definite  
relationship between the electron and proton spectra from a same source,
so they proposed the energy region and flux of the second possible excess of electron from the proton spectra. 
However, for the GC process $p+p(A)\to \pi^\pm + \pi^0+ p + \overline{p}+\text{other}$ and $\pi^0 \to e^+ + e^-$, they only considered the final secondary particles electrons and protons, and didn't consider the secondary particles gamma rays from the decay of $\pi^0$, as well as  the processes where $\pi^\pm$ and $\mu^\pm$ decay into neutrinos. So, in this work we explore the possible gamma-ray and neutrino spectra relating to the second excess of electron based on the connections between these  spectra proposed by the GC model. The first possible gamma-ray spectra relating to the first excess of electron has been studied in works\cite{Zhu:2017bvp}.

This paper is organized as follows. Section 2 provides a brief introduction to the GC model and derives the spectrum equation for the cosmic ray neutrino. In the section 3, based on the cosmic ray particle equation given by the GC model, we fit the source LHAASO J2108-5157 and establish the connection between the $\gamma$ spectrum and the neutrino spectrum. Finally, in the section 4 we give a discussion and summary.

\section{The GC Model}
\label{sec:2}
As we know the secondary CR particles may originate from hadronic processes, such as $p+p(A)\to \pi^\pm + \pi^0 + p+\overline{p}+ \text{other}$, as well as subsequent processes like $\pi^0\to 2\gamma$ and the decay of $\pi^\pm$. Define $N_\pi (E_{p-p(A) },E_\pi)$ as the number of mesons with energy $E_\pi$ in  $p-p(A)$ collisions, where $E_{p-p(A)}$ is the energy of the incident proton in the rest frame of the target proton. Due to the non-perturbative hadronization process of $p-p(A)$ collisions, it's very hard to calculate the distribution of $\pi$ mesons. To simplify the calculation, in GC model, we only consider mesons as secondary particles, as their diversity is much greater than that of other particles in high-energy collisions. Typically, these $\pi$ mesons have relatively low kinetic energy (or momentum) in the center-of-mass (CM) system, especially in the central region of the rapidity distribution. At a given interaction energy, the maximum value of $N_\pi$  can only occur when almost all available kinetic energy in the CM system is used to produce $\pi$ mesons. We assume that huge  number of gluons are produced in the central region due to the GC effect which leads to the maximum value of $N_\pi$. It's worth noting that this assumption is made for computational simplification purposes and is not a requirement of the GC condition. We will demonstrate the computational simplification brought by this assumption, and it does not fundamentally alter the characteristics of GC. Considering the relativistic invariants and energy conservation, the following equation holds \cite{Zhu:2017bvp}:
\begin{equation}\label{eq:2.1}
	\begin{aligned}
		(2m_p^2+2E_{p-p(A)}m_p)^{1/2} &= E_{p1}^* + E_{p2}^* + N_{\pi} m_{\pi},\\
		E_{p-p(A)} + m_p &= m_p \gamma_1 + m_p \gamma_2 + N_{\pi} m_{\pi} \gamma.
	\end{aligned}
\end{equation}
Here, $E_{(p_i)}^*$ is the energy of the leading protons in the CM system, $\gamma_i$ is the Lorentz factor for the corresponding particles. Considering the parameter $K \sim 0.5$ \cite{Gaisser:1990vg}, the eq.\ref{eq:2.1} can be simplified to:
\begin{equation}\label{eq:2.2}
	\begin{aligned}
		E_{p1}^* + E_{p2}^* &= \left(\frac{1}{K} - 1\right) N_{\pi} m_{\pi},\\ 
		m_p \gamma_1 + m_p \gamma_2 &= \left(\frac{1}{K} - 1\right) N_{\pi} m_{\pi} \gamma.
	\end{aligned}
\end{equation}
For $p-p(A)$ collisions, the solution for $N_{\pi}(E_{p-p(A)}, E_{\pi})$ can be expressed as:
\begin{equation}\label{eq:2.3}
	\begin{aligned}
		\ln N_{\pi} &= 0.5 \ln E_{p-p(A)} + a, \\
		\ln N_{\pi} &= \ln E_{\pi} + b,
	\end{aligned}
\end{equation}
where
\begin{equation}\label{eq:2.4}
	\begin{aligned}
		a &\equiv 0.5 \ln (2m_p) - \ln m_{\pi} + \ln K, \\
		b &\equiv \ln (2m_p) - 2 \ln m_{\pi} + \ln K.
	\end{aligned}
\end{equation}
In these equations, $E_{\pi} \in [E_{\pi}^{GC}, E_{\pi}^{\text{max}}]$. Eq.\ref{eq:2.3} establishes a direct relationship between $N_{\pi}$, $E_{p-p(A)}$ and $E_{\pi}^{GC}$, thereby facilitating the derivation of the spectrum of gamma ray resulting from GC model.

Generally, the spectrum of gamma ray is given by:
\begin{equation}\label{eq:2.5}
	\Phi_{\gamma}(E_{\gamma}) = \Phi_{\gamma}^0(E_{\gamma}) + \Phi_{\gamma}^{GC}(E_{\gamma}).
\end{equation}
Here, $\Phi_{\gamma}^0(E_{\gamma})$ denotes the background gamma spectrum. In the GC model, the local contribution of gamma ray is:
\begin{equation}\label{eq:2.6}
	\begin{aligned}
		\Phi_{\gamma}^{GC}(E_{\gamma}) &= C_{p-p(A)} \left(\frac{E_{\gamma}}{E_0}\right)^{-\beta_{\gamma}} \int_{E_{\pi}^{\text{min}}}^{E_{\pi}^{\text{max}}} \mathrm{d}E_{\pi} \left(\frac{E_{p-p(A)}}{E_{p-p(A)}^{GC}}\right)^{-\beta_p}\\
		&\times N_{\pi}(E_{p-p(A)}, E_{\pi}) \left(\frac{\mathrm{d}\omega_{\pi-\gamma}(E_{\pi},E_{\gamma})}{\mathrm{d}E_{\gamma}}\right).
	\end{aligned}
\end{equation}
In this equation, $\beta_\gamma$ signifies the propagation losses of gamma rays, and $\beta_p$ is related to the proton acceleration mechanism. The parameter $C_{p-p(A)}$ combines the kinematic factors and the dimensionality of the proton spectrum with the branching ratio of the $\pi^0 \to 2\gamma$ process. The normalized spectrum for $\pi^0 \to 2\gamma$ is:
\begin{equation}\label{eq:2.7}
	\frac{\mathrm{d}\omega_{\pi-\gamma}(E_{\pi}, E_{\gamma})}{\mathrm{d}E_{\gamma}} = \frac{2}{\beta_{\pi} E_{\pi}} H\left[E_{\gamma}; \frac{1}{2} E_{\pi} (1 - \beta_{\pi}), \frac{1}{2} E_{\pi} (1 + \beta_{\pi})\right],
\end{equation}
where $H(x;a,b)$ represents the Heaviside function, which equals 1 when $a \leq x \leq b$ and 0 otherwise. By substituting eqs.\ref{eq:2.3}, \ref{eq:2.4} and \ref{eq:2.7} into eq.\ref{eq:2.6}, the GC-characteristic spectrum of gamma ray is obtained:
\begin{equation}\label{eq:2.8}
	\begin{aligned}
		\Phi_{\gamma}^{GC}(E_{\gamma}) &= C_{p-p(A)} \left(\frac{E_{\gamma}}{E_{\pi}^{\text{GC}}}\right)^{-\beta_{\gamma}} \int_{E_{\pi}^{\text{GC}} \text{ or } E_{\gamma}}^{E_{\pi}^{\text{GC,max}}} \mathrm{d}E_{\pi} \left(\frac{E_{p-p(A)}}{E_{p-p(A)}^{\text{GC}}}\right)^{-\beta_p} \\
		&\times N_{\pi}(E_{p-p(A)}, E_{\pi}) \frac{2}{\beta_{\pi} E_{\pi}}.
	\end{aligned}
\end{equation}
If $E_\gamma \leq E_\pi^\mathrm{GC}$ (or $E_\gamma>E_\pi^\mathrm{GC}$), the lower limit of the integral is $E_\pi^\mathrm{GC}$ (or $E_\gamma$). Finally, integrating this expression gives the parameterized form of gamma ray\cite{Zhu:2020gys,Ruan:2020whn}:
\begin{equation}\label{eq:2.9}
	\Phi_{\gamma}^{\mathrm{GC}}(E_{\gamma}) =
	\begin{cases} \displaystyle
		\frac{50C_\gamma}{2\beta_p - 1} E_{\pi}^{\mathrm{GC}} \left(\frac{E_{\gamma}}{E_{\pi}^{\mathrm{GC}}}\right)^{-\beta_{\gamma}} & \text{if } E_{\gamma} \leq E_{\pi}^{\mathrm{GC}}, \\ \displaystyle
		\frac{50C_\gamma}{2\beta_p - 1} E_{\pi}^{\mathrm{GC}} \left(\frac{E_{\gamma}}{E_{\pi}^{\mathrm{GC}}}\right)^{-\beta_{\gamma} - 2\beta_p + 1} & \text{if } E_{\gamma} > E_{\pi}^{\mathrm{GC}}.
	\end{cases}
\end{equation}
This parameterized form is a power-law function with a sharp break. The pure power-law form for $E_{\gamma} \leq E_{\pi}^{\mathrm{GC}}$ arises from the fixed lower limit of integration in eq.\ref{eq:2.8}, around $E_\pi^\mathrm{GC}$. The characteristics of $\Phi_\gamma^\mathrm{GC}$  mentioned above are directly caused by the GC effect and are termed GC-characteristics. They differ from all other known smooth radiation spectra. In eq.\ref{eq:2.9}, the second power-law function for $E_\gamma>E_\pi^\mathrm{GC}$ results from the simplified assumption in eq.\ref{eq:2.1}, which assumes that all available kinetic energy in the central region is used to produce $\pi$ mesons. It's important to note that if this simplification is modified, the power-law for $E_\gamma> E_\pi^\mathrm{GC}$ under the integral conditions becomes variable. This parameterized form can be compared with experimental data to test the validity of the simplification. Nevertheless, it does not alter the GC-characteristics mentioned above.

Considering the processes $\pi^{\pm} \rightarrow \mu^{\pm} + \nu_{\mu} (\bar{\nu}_{\mu})$ and $\mu^{\pm} \rightarrow e^{\pm} + \nu_e (\bar{\nu}_e) + \bar{\nu}_{\mu} (\nu_{\mu})$ which produce neutrinos and electrons, the GC spectrum for electrons is given by \cite{Zhu:2017bvp}:
\begin{equation}\label{eq:2.10}
	\begin{aligned}
		\Phi_{e}^{\mathrm{GC}}(E_{e}) &= C_{e} \left(\frac{E_{e}}{E_{\pi}^{\mathrm{GC}}}\right)^{-\beta_{e}} \int \mathrm{d}E_{\mu} \int_{2.5E_{e} \text{ or } E_{\pi}^{\mathrm{GC}}}^{E_{\pi}^{max}} \frac{\mathrm{d}E_{\pi}}{E_{\pi}} \left(\frac{E_{p-p(A)}}{E_{p-p(A)}^{\mathrm{GC}}}\right)^{-\beta_p}\\ 
		&\times N_{\pi^{\pm}}(E_{p-p(A)}, E_{\pi}) \left(\frac{\mathrm{d}\omega_{\pi-\mu}(E_{\pi},E_{\mu})}{\mathrm{d}E_{\mu}}\right) \left(\frac{\mathrm{d}\omega_{\mu-e}(E_{\mu},E_{e})}{\mathrm{d}E_{e}}\right).
	\end{aligned}
\end{equation}
Similarly, the GC spectrum for neutrinos can be derived as:
\begin{equation}\label{eq:2.11}
	\begin{aligned}
		\Phi_{\nu}^{\mathrm{GC}}(E_{\nu}) &= C_{\nu} \left(\frac{E_{\nu}}{E_{\pi}^{\mathrm{GC}}}\right)^{-\beta_{\nu}} \int \mathrm{d}E_{\mu} \int_{E_{\pi}^{\mathrm{GC}} \text{ or } E_{\nu}}^{E_{\pi}^{max}} \frac{\mathrm{d}E_{\pi}}{E_{\pi}} \left(\frac{E_{p-p(A)}}{E_{p-p(A)}^{\mathrm{GC}}}\right)^{-\beta_p}\\ 
		&\times N_{\pi^{\pm}}(E_{p-p(A)}, E_{\pi}) \left(\frac{\mathrm{d}\omega_{\pi-\mu}(E_{\pi},E_{\mu})}{\mathrm{d}E_{\mu}}\right) \left(1 + \frac{\mathrm{d}\omega_{\mu-\nu}(E_{\mu},E_{\nu})}{\mathrm{d}E_{\nu}}\right)\\
		&= C_{\nu} E_{\pi}^{\mathrm{GC}}\left(\frac{E_{\nu}}{E_{\pi}^{\mathrm{GC}}}\right)^{-\beta_{\nu}} \\
		&\times 
	\begin{cases}
		(\frac{-31}{2\beta_p - 1} + \frac{240}{\beta_p} \frac{E_{\nu}}{E_{\pi}^{\mathrm{GC}}}) & \text{if } E_{\nu} \leq E_{\pi}^{\mathrm{GC}}, \\ 
	   \left(\frac{-31}{2\beta_p - 1} + \frac{240}{\beta_p} \right) \left(\frac{E_{\nu}}{E_{\pi}^{\mathrm{GC}}}\right)^{-2\beta_p + 1} & \text{if } E_{\nu} > E_{\pi}^{\mathrm{GC}}.
	\end{cases}
	\end{aligned}
\end{equation}
 The normalization spectra in the above integration are given by:
\begin{equation}\label{eq:2.12}
	\begin{aligned}
		\frac{\mathrm{d}\omega_{\pi-\mu}(E_{\pi}, E_{\mu})}{\mathrm{d}E_{\mu}} &= \delta(E_{\mu} - 0.8E_{\pi}),\\
		\frac{\mathrm{d}\omega_{\mu-\nu}(E_{\mu}, E_{\nu})}{\mathrm{d}E_{\nu}} &= 16\left(1 - \frac{E_{\nu}}{E_{\mu}}\right)^2 \left(\frac{2E_{\nu}}{E_{\mu}} - 0.5\right).
	\end{aligned}
\end{equation}
It is important to note that the lower limit of the integral in eq.\ref{eq:2.11} is  $\max(E_{\pi}^{\mathrm{GC}}, E_{\nu})$, and it also shows a sharp break around $E_{\nu} = E_{\pi}^{\mathrm{GC}}$.
The parameters $\beta_e$ and $\beta_{\nu}$ in eqs.\ref{eq:2.10} and \ref{eq:2.11} indicate the propagation losses of electrons and neutrinos.

\section{Predictions}
\label{sec:3}
In our previous work \cite{Wu:2023fbw} , the second possible electron excess was predicted. According to the GC model, the same GC source is expected to simultaneously produce corresponding gamma-ray and neutrino emissions. In work \cite{Wu:2023fbw}, potential sources were predicted with a GC threshold of $E_\pi^\mathrm{GC}=24.0\mathrm{~TeV}$, and the corresponding parameters were chosen as  $\beta_p=1.8$, $\beta_e=0.2$, and $\beta_\gamma=0.3$. This section explores the possible simultaneous production of gamma-ray and neutrino spectra based on the GC model.

As in work \cite{Wu:2023fbw} and eqs.\ref{eq:2.9} and \ref{eq:2.10}, all the three spectra (electron, gamma-ray, neutrino) contain the parameter $\beta_p$ for proton propagation energy loss. The electron and gamma-ray spectra contain the parameter $\beta_\gamma$ for gamma-ray propagation energy loss, and the neutrino spectrum includes the parameter $\beta_\nu$ for neutrino propagation energy loss.
According to the GC model, the cosmic rays come from a same GC source should share the same propagation parameters. So, as in \cite{Wu:2023fbw}, relating to the second excess of electron, we take $\beta_p=1.8$, $\beta_e=0.2$, and $\beta_\gamma=0.3$. 
It is known from the literature \cite{Liu:2019gxw,Wu:2023fbw} that the GC threshold for gamma rays and neutrinos is the same as that for electrons, $E_{\gamma}^{\mathrm{GC}} = E_{\nu}^{\mathrm{GC}} = E_{e}^{\mathrm{GC}} = 24 \, \text{TeV}$.

\begin{figure}[htbp]
	\centering 
	{\includegraphics[width=0.6\textwidth]{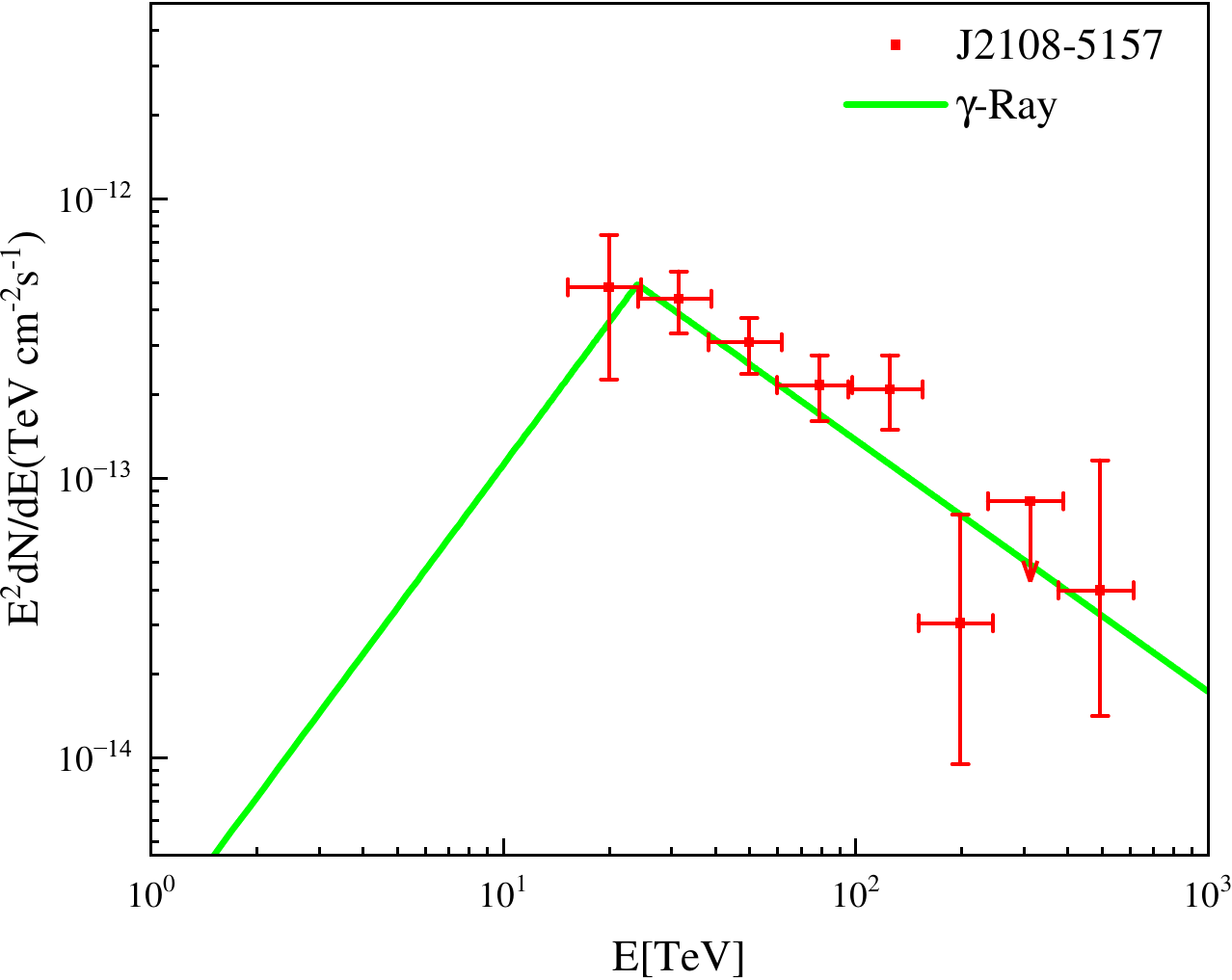}}
	\caption{Gamma-ray Spectrum. The green solid line represents the GC gamma-ray spectrum at the GC threshold $E_\gamma^\mathrm{GC}=24.0 \mathrm{~TeV}$. The red dots indicate the observed data of the source LHASSO J2108+5157 \cite{LHAASO:2021quh}.}
	\label{fig.1}
\end{figure}

According to eq.\ref{eq:2.9}, we can now draw the gamma-ray spectra except that the parameter $C_\gamma$ is unkown. 
Searched for all the possible
gamma-ray spectra, it was found that the source J2108+5157 detected by LHASSO exhibits the GC characteristics, with spectral parameters matching the aforementioned values. Based on the observed data of LHASSO J2108+5157, the parameter $C_\gamma=3.71\times10^{41}$ was determined in eq.\ref{eq:2.9}, and the corresponding spectrum is shown in Fig.\ref{fig.1}.

\begin{figure}[htbp]
	\centering 
	{\includegraphics[width=0.6\textwidth]{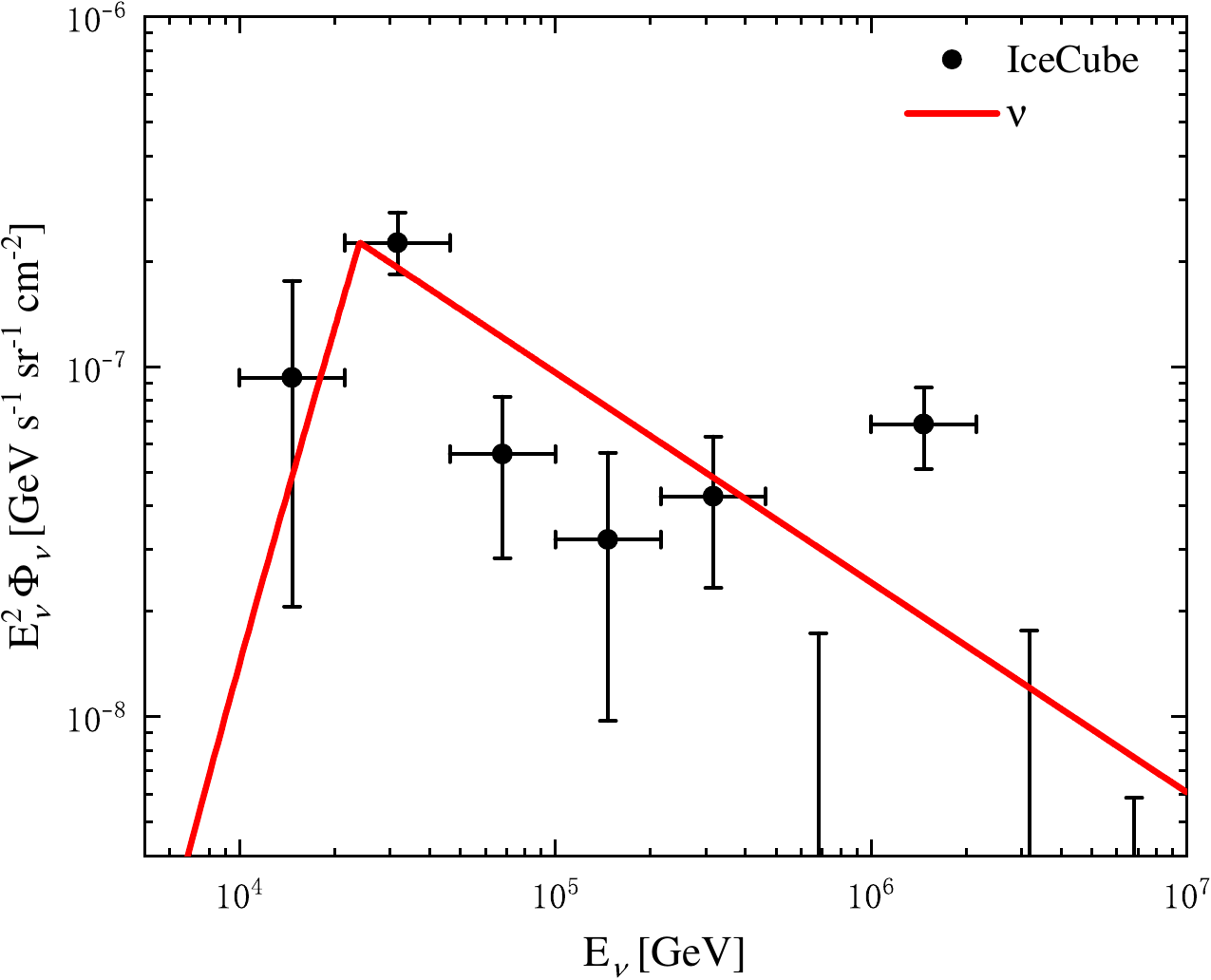}}
	\caption{Neutrino spectrum. The black dots represent the observational data from the IceCube experiment \cite{IceCube:2015gsk}.}
	\label{fig.2}
\end{figure}

For the neutrino spectra as eq.\ref{eq:2.11}, the parameters $C_\nu$ and  $\beta_\nu$ are unkown. We take  $\beta_\nu=0$
since neutrinos hardly lose their energy during the transmission. The parameters $C_\nu$ should be fitted with experiments,
while there is very limited experimental data available for neutrino observations. Fig.\ref{fig.2} shows the observation data from the IceCube \cite{IceCube:2015gsk} experiment. Using eq.\ref{eq:2.11}, with $\beta_p = 1.8$, $\beta_\nu=0$, and $C_\nu=1.42\times 10^{-22}$ ,  the spectrum is shown in Fig.\ref{fig.2}. Considering the large errors of the observation data, our results appear to be consistent with the experiment data almostly.

From Figs.\ref{fig.1} and \ref{fig.2}, it can be seen that both the gamma-ray and neutrino spectrum curves exhibit distinct GC features, with the typical GC break occurring at around 24.0 TeV.

\section{Discussion and summary}
\label{sec:4}
During intense astrophysical processes, the hadron collidings often happen, cosmic rays, such as
electrons (positrons), protons (antiprotons), gamma rays, and neutrinos are often generated concurrently. 
However, due to complecated interaction mechanisms, it's hard to make a clear correlations among different type of spectra. 
In our work, GC phenomena may happen during ultra-high-energy collisions.
 According to this model, cosmic rays originate from one certain GC process have definite correlations between different types of spectra. So, in our previous work\cite{Wu:2023fbw}, a second electron excess was infered which originated from the GC process of a local source, the possible shape and amplitude of its spectra were predicted based on proton data. The primary focus of this paper is the prediction of gamma-ray and neutrino spectra potentially generated by this local source.

According to work\cite{Wu:2023fbw} and eqs.\ref{eq:2.9} and \ref{eq:2.11}, it is evident that for the electron, gamma-ray and neutrino spectra, there is a common parameter $ \beta_p$, and their GC thresholds are also identical. Based on this characteristic, LHASSO J2108+5157 was identified as the source that best matches the GC-characterization, $\beta_p$ and $E_\gamma^\mathrm{GC}$. Using the observational data, its amplitude was determined. For the neutrino spectrum, which has only several observational data, the amplitude  was determined using the observation data from the IceCube experiment.

Considering that gamma-rays and neutrinos propagate in straight lines in cosmic space, it is possible to determine the local sources that contributes to the second electron excess by examining the gamma sources. However, since LHASSO did not detect the specific data (years and distance) for the source J2108+5157, it cannot be confirmed whether this source  contributes to the second electron excess or not. Similarly, the observational data for neutrinos only provide flux data while without specifying the astrophysical sources, making it impossible to verify the connection between their sources. Ultimately, more precise experiments are anticipated to enrich the data of electrons, gamma- rays and neutrinos, as well as parameters of radiation sources, to validate our model.

\acknowledgments

This work is supported by the National Natural Science Foundation of China (No.11851303).



\bibliographystyle{JHEP} 
\bibliography{ref} 



\end{document}